\documentclass[english,prl,aps,reprint,longbibliography]{revtex4-1}
\usepackage[T1]{fontenc}
\usepackage[latin9]{inputenc}
\usepackage{textcomp}
\usepackage{amsmath}
\usepackage{amssymb}
\usepackage{graphicx}

\makeatletter

\newcommand*\LyXThinSpace{\,\hspace{0pt}}

\usepackage{babel}

\makeatother

\usepackage{babel}
\begin{document}
\title{Fractional Electromagnetic Response in Three-Dimensional Chiral Anomalous
Semimetal}
\author{Huan-Wen Wang, Bo Fu, Jin-Yu Zou, Zi-Ang Hu, Shun-Qing Shen}
\email{sshen@hku.hk}

\affiliation{Department of Physics, The University of Hong Kong, Pokfulam Road,
Hong Kong, China}
\date{\today}
\begin{abstract}
The magnetoelectric coupling of electrons in a three-dimensional solid
can be effectively described by axion electrodynamics. Here we report
the discovery of the fractional magnetoelectric effect in chiral anomalous
semimetals of the three-dimensional massless Wilson fermions, which
have linear dispersions at the energy crossing point, and break the
chiral symmetry at generic momenta. In the presence of electric and
magnetic fields, the time-reversal and parity symmetry breaking give
rise to the quarter-quantized topological magnetoelectric effect,
which is directly related to the winding number 1/2 of the band structure,
and is only one half of that for topological insulators. The fractional
electromagnetic response can be revealed by the surface Hall conductance
and extracted from the measurement of the topological Kerr and Faraday
rotation. The transition-metal pentatelluride $\mathrm{ZrTe_{5}}$
with a strain-tunable band gap provides a potential platform to test
the effect experimentally.
\end{abstract}
\maketitle

\section{Introduction}

The electromagnetic response of a three-dimensional insulator is described
by an effective axion action $S_{\theta}=\frac{\theta}{2\pi}\frac{e^{2}}{h}\int d^{3}xdt\mathbf{E}\cdot\mathbf{B}$
in additional to the conventional Maxwell action \citep{Wilczek1987axiondynamics}.
Here $\mathbf{E}$ and $\mathbf{B}$ are the electromagnetic fields
inside the insulator, and $\theta$ is the axion field. All time-reversal
invariant insulators fall into two distinct classes described by either
$\theta=0$ (for trivial insulators) or $\theta=\pm\pi$ (for topological
insulators) \citep{Qi2008QFT,Qi2011rmp}. The time reversal invaraince
of topological insulator makes the $\theta$-term quantized and leads
to the topological magnetoelectric effect in units of fine structure
constant $\alpha=e^{2}/\hbar c$ \citep{Spaldin2005ME,Fiebig2005ME,Essin2009OMP}.
In general, the axion field $\theta$ can be evaluated from the zero-field
non-Abelian Berry connection of the band structure as $\theta=\frac{1}{4\pi^{2}}\int_{BZ}d^{3}k\epsilon^{lmn}\mathrm{Tr}[\mathcal{A}_{l}\partial_{m}\mathcal{A}_{n}-\frac{2i}{3}\mathcal{A}_{l}\mathcal{A}_{m}\mathcal{A}_{n}]$,
where $\mathcal{A}_{l}^{\mu\nu}=i\langle u_{\mu}|\partial_{l}u_{\nu}\rangle$
is the Berry connection defined from the Bloch function of occupied
band $\mu$ and $\nu$, $\epsilon^{lmn}$ is the Levi-Civita symbol,
the indices $l,m,n$ run over $1,2,3$ and $\partial_{l}=\partial/\partial k_{l}$
\citep{Qi2008QFT,Essin2009OMP}. This Chern-Simons contribution to
the magnetization remains parallel to electric field for arbitrary
orientations of the external field relative to the crystal axes and
vanishes in less than three dimensions. With an open boundary condition,
the topological magnetoelectric effect corresponds to the half-quantized
surface Hall conductance subjected to a symmetry-broken Zeeman field
\citep{Qi2008QFT,Chu2011prb}. Now a question on the validity of the
axion action arises when the band gap of the system closes. Furthermore,
even if the action exists, what is the value of the action $\theta$?
Does the surface Hall effect still hold? These are the key issues
that we want to address in the present work.

Massless Wilson fermions arise from the lattice regularization for
Dirac fermions in the lattice gauge theory, which possess the linear
dispersion near the energy crossing point, but breaks the chiral symmetry
at higher energy \citep{Wilson1977book,Nielsen1981PLB,Rothe1998book}.
The massless Wilson fermions may constitute a novel type of topological
quantum semimetal, named chiral anomalous semimetal (CAS). This topological
state is not prohibited by the Nielsen-Ninomiya theorem and avoid
the fermion doubling problem \citep{Nielsen1981PLB} and is characterized
by a one-half topological invariant \citep{FuboQAS,Zou2022arxiv}.
This is opposite to all existing topological states, such as quantum
Hall effect, topological insulators and topological superconductors,
which are always characterized by integer topological invariants,
i.e., $\mathbb{Z}$ or $\mathbb{Z}_{2}$ index \citep{vonKlitzing-80prl,Klitzing-20nrp,Moore2010nature,Hasan2010rmp,Qi2011rmp,armitage2018rmp,SQS,Tokura-18nrp}.

In this paper, we derived the continuity equation for chiral current,
which restores the chiral anomaly near the band crossing point. We
further present the derivation of the $\theta$ field for CAS in the
presence of electromagnetic field. The value of $\theta$ is found
to be $\pm\pi/2$ subjected to the time-reversal and parity symmetry
breaking, leading to a quarter-quantized topological magnetoelectric
effect. The $\theta$ field and spatial distribution of magneto-electric
polarization can be revealed by the surface Hall conductance, although
there is no surface state, and extracted from the measurement of the
topological Kerr and Faraday rotation at reflectivity minimum and
maximum, which can provide a substantial evidence for the existence
of CAS in solids. Lastly, the interaction, strain and temperature
tunable axion field have been discussed near the topological phase
transition point in transition-metal pentatelluride $\mathrm{ZrTe_{5}}$.

This paper is organized as follows. In Sec. II, we present the effective
model Hamiltonian for transition-metal pentatelluride $\mathrm{ZrTe}_{5}$.
In Sec. III, we discuss the quantum anomaly for CAS. In Sec. IV, the
fractional magnetoelectric effect has been discussed for the time-reversal
symmetry broken CAS. In Sec. V, the surface hall effect and surface
current are addressed in the slab geometry. In Sec. VI, the spin density
wave order has been derived to realize the symmetry breaking term
near the phase transition point.

\section{Model Hamiltonian}

In electronic system, the CAS states can be realized near the topological
phase transition points. The first principles calculation and the
angle-resolved photoemission spectroscopy (ARPES) measurement demonstrate
that the band structures of transition-metal pentatelluride $\mathrm{ZrTe}_{5}$
is the prototype of massive Dirac material \citep{weng2014prx,Chen2015prl,Li16natphys,mutch2019evidence,zhang2021natcomm},
which is very close to the critical points. In the basis $(|\mathrm{Te_{1}}p_{y}\uparrow\rangle,|\mathrm{Te_{1}}p_{y}\downarrow\rangle,|\mathrm{Te_{2}}p_{y}\uparrow\rangle,|\mathrm{Te_{2}}p_{y}\downarrow\rangle)$,
with $\mathrm{Te_{1/2}}p_{y}$ being the $p_{y}$ orbital of the two
$\mathrm{Te}$ atoms in the unit cell, the low energy $\mathbf{k}\cdot\mathbf{p}$
Hamiltonian can be described by the following modified Dirac equation
\citep{weng2014prx,Chen2015prl} 
\begin{equation}
H_{0}(k)=(m-b\hbar^{2}k^{2})\tau_{1}+v\hbar(k_{x}\tau_{3}\sigma_{3}+k_{y}\tau_{2}+k_{z}\tau_{3}\sigma_{1}),\label{eq:model}
\end{equation}
where $v$ is the effective velocity, $k_{i}$ with $i=x,y,z$ are
the wavevector operators. $\mathbf{\sigma}=\{\sigma_{1},\sigma_{2},\sigma_{3}\}$
and $\mathbf{\tau}=\{\tau_{1},\tau_{2},\tau_{3}\}$ are the Pauli
matrices acting on the spin and orbital space, respectively. $m-b\hbar^{2}k^{2}$
is the momentum-dependent Dirac mass. Here we focus on the case of
semi-metallic state ($m=0$). When $b=0$ and $m=0$, Eq. (\ref{eq:model})
describes the massless Dirac fermion with linear dispersion, and it
has the chiral symmetry with chiral operator $\gamma_{5}=\tau_{2}\sigma_{2}$,
which is forbidden in a lattice case as required by the Nielsen-Ninomiya
theorem. When $b\ne0$ and $m=0$, Eq. (\ref{eq:model}) describes
the CAS.

Under pressure, $\mathrm{ZrTe_{5}}$ will undergo a topological phase
transition which is accompanied with the closing and reopening of
the band gap.The pressure does not couple with the spin degrees of
freedom and preserves time-reversal symmetry. We then introduce the
electron-strain coupling around $\Gamma$ point based on the time
reversal symmetry:
\begin{equation}
H_{strain}=(u_{xx}\xi_{xx}+u_{yy}\xi_{yy}+u_{zz}\xi_{zz})\tau_{1}\label{eq:strain}
\end{equation}
which is the only symmetry allowed momentum independent term. The
strain (\ref{eq:strain}) induced by the axial stress does not break
the $D_{2h}$ point group symmetry. $\xi_{ii}$ are the material-dependent
coupling constants between the low- energy electrons and the strain
tensor $u_{ii}=\partial_{i}u_{i}$ with $\boldsymbol{u}(\boldsymbol{x})=(u_{1},u_{2},u_{3})$
being the displacement field at $\boldsymbol{x}$. Thus, stretching
the crystal along the $z$ direction can be represented by a displacement
field $\boldsymbol{u}=(0,0,\alpha z)$ with the only nonzero strain
tensor $u_{33}=\alpha$ where $\alpha=\Delta L/L$ measures the elongation
of the crystal. In the presence of external strain, $m\to m+\alpha\xi_{zz}$,
the Dirac mass varies linearly with the strain. Such a strained-dependent
Dirac mass has been tested by several experiments \citep{mutch2019evidence,zhang2021natcomm}.
Hence, one can use strain to tune the mass of $\mathrm{ZrTe_{5}}$
to zero, thus forming CAS.

\section{Continuity equation for chirality}

The presence of the $b\hbar^{2}k^{2}\tau_{1}$ term in Eq. (\ref{eq:model})
breaks the chiral symmetry explicitly, $[\gamma_{5},\tau_{2}\sigma_{2}]\ne0$.
Following the Jackiw-Johnson approach to the chiral anomaly \citep{Jackiw1969pr,wang2021prb},
we can derive the continuity equation for the chiral current as (see
Appendix \ref{sec:Continuity-equation-for} for details) 
\begin{equation}
\nabla\cdot\mathbf{j}_{5}+\frac{\partial\rho_{5}}{\partial t}=\frac{1}{\sqrt{1+(b\hbar k_{F}/v)^{2}}}\frac{e^{2}}{2\pi^{2}\hbar^{2}}\mathbf{E}\cdot\mathbf{B}\label{eq:chiral anomaly}
\end{equation}
where $\mathbf{j}_{5}$ and $\rho_{5}$ are the chiral current and
chiral density, respectively, $k_{F}$ is the fermi wavevector, and
$\mathbf{E}$ and $\mathbf{B}$ are the electric field and magnetic
field, respectively. In the limit of $b\to0$ or $k_{F}\to0$, the
energy dispersion is almost linear in momentum. The chiral symmetry
is restored in this case, and the chirality should be conserved. However,
the left hand of equation does not vanish in the presence of an electromagnetic
field. In fact we derive successfully the continuity equation for
the chiral anomaly \citep{Adler1969pr,Bell1969PCAC}. The presence
of the chiral anomaly in the system is the reason we name it as CAS.
Different from the ideal linearized Dirac fermion, the $\mathbf{E}\cdot\mathbf{B}$
term is from the symmetry breaking term $b\hbar^{2}k^{2}\tau_{1}$
in CAS instead of the spontaneous chiral symmetry breaking from the
infinite Dirac sea. Recently, the effect is shown to be closely related
to the helical symmetry breaking in the presence of an electric field
\citep{wang2021prb}. Besides, the $b$ term also plays an important
role in parity anomaly. In 1+2 dimensional system, the $b$ term allows
the existence of single Dirac cone and half-quantized Hall conductance
on the time-reversal symmetry breaking lattice \citep{FuboQAS,Zou2022arxiv}.

\section{Fractional orbital magnetoelectric polarization}

Although the chiral symmetry is broken in high energy, the system
still possesses an additional sublattice symmetry as $(\tau_{3}\sigma_{2})H_{0}(k)(\tau_{3}\sigma_{2})=-H_{0}(k)$.
Under the rotation transformation in the orbital and spin space $\mathcal{U}=\exp(i\frac{\pi}{3}\tau\cdot\hat{n})\exp\left(i\frac{\pi}{4}(\tau_{0}+\tau_{1})\sigma_{2}\right)$,
where $\hat{n}$ is the unit vector along the direction $(1,1,1)$,
$H_{0}$ is brought into an off-diagonal form as $\begin{pmatrix}0 & q\\
q^{\dagger} & 0
\end{pmatrix}$ with $q=v\hbar\mathbf{k\cdot\sigma}^{\prime}+ib\hbar^{2}(\mathbf{k\cdot\sigma}^{\prime})^{2}$
and $\sigma^{\prime}=\{\sigma_{1},\sigma_{2},-\sigma_{3}\}$. Then,
the topological property of $H_{0}$ can be characterized by the three-dimensional
winding number \citep{Schnyder2008classification,Ryu2010tenfold}
\begin{align*}
w_{3D} & =\int d^{3}\mathbf{k}\varpi_{3D},\\
\varpi_{3D} & =\frac{\epsilon^{lmn}}{24\pi^{2}}\mathrm{tr}[(\hat{q}^{-1}\partial_{l}\hat{q})(\hat{q}^{-1}\partial_{m}\hat{q})(\hat{q}^{-1}\partial_{n}\hat{q})]
\end{align*}
 where $\varpi_{3D}$ is the winding number density and $\hat{q}=q/|q|$.
After some tedious algebra, one finds 
\begin{equation}
w_{3D}=-\frac{1}{2}\mathrm{sgn}(b).\label{eq:theta_m2to0}
\end{equation}
Hence, the CAS is strikingly distinct from the existing topological
phases with integer topological invariants. In general, the half-quantized
winding number can be ascribed to the chiral symmetry around the energy
crossing point and is classified by the relative homotopy group \citep{FuboQAS}.

After carefully examing numerically and analytically, we find no evidence
to support the existence of the surface states around the CAS in contrast
to the topological insulators. As shown in Ref. \citep{Ryu2010tenfold},
the integer winding number is related to the half-quantized orbital
magnetoelectric polarization in the presence of sublattice symmetry
for a gapped system. Then, one natural question raises here: does
the one-half winding number indicate a quarter-quantized orbital magnetoelectric
polarization in the absence of the surface states? The answer is yes
as shown blow.

To explore the magnetoelectric effect, we introduce a symmetry breaking
term $m_{2}\tau_{3}\sigma_{2}$, which destroys both the time-reversal
and spatial inversion symmetry but preserves their combination \citep{Li2016np,sekine2014jpsj},
into the original Hamiltonian. The term acts as the spin density wave
order and will be discussed later. The topological $\theta$ term
can be obtained by integrating the Chern-Simons three form over the
Brilliun zone. Unlike the winding number, the $\theta$ term is defined
without assuming the sublattice symmetry and can be used for system
with symmetry breaking term $m_{2}$. For sublattice symmetric Hamiltonian
($m_{2}=0$), the eigen wavefunction can be constructed as $|u_{a}^{s}\rangle=\frac{1}{\sqrt{2}}[n_{a},s\hat{q}^{\dagger}n_{a}]^{T}$
or for a different choice of gauge $|v_{a}^{s}\rangle=\frac{1}{\sqrt{2}}[\hat{q}n_{a},sn_{a}]^{T}$,
where $s=\pm$ denotes the conduction and valence bands, $a=1,2$
represent the degenerate two states and $(n_{a})_{b}=\delta_{ab}$
are $k-$independent orthonormal vectors. The presence of $m_{2}$
will mix the states with different subscripts, and the eigen wavefunction
$|\phi_{a}^{s}\rangle$ can be expressed as a linear combination of
these basis. In order to evaluate the Berry connection, the wave function
should be constructed without any singularity. By choosing a gauge
properly, the well-defined wavefunctions are 
\begin{align*}
|\phi_{a}^{-}\rangle= & \begin{cases}
\cos\frac{\varphi_{q}}{2}|u_{a}^{+}\rangle+\sin\frac{\varphi_{q}}{2}|u_{a}^{-}\rangle & m_{2}<0\\
\cos\frac{\varphi_{q}}{2}|v_{a}^{+}\rangle-\sin\frac{\varphi_{q}}{2}|v_{a}^{-}\rangle & m_{2}>0
\end{cases}
\end{align*}
with $\sin\varphi_{q}=\frac{|m_{2}|}{\sqrt{m_{2}^{2}+qq^{\dagger}}}$
and $\cos\varphi_{q}=\frac{\sqrt{qq^{\dagger}}}{\sqrt{m_{2}^{2}+qq^{\dagger}}}$.
From which, we can obtain the $\theta$ term as 
\begin{equation}
\frac{\theta}{2\pi}=\frac{1}{4}\mathrm{sgn}(m_{2})\int d^{3}\mathbf{k}(2-3\sin\varphi_{q}+\sin^{3}\varphi_{q})\varpi_{3D}.
\end{equation}
As taking $m_{2}\to0$, $\sin\varphi_{q}\to0$, we have 
\begin{equation}
\lim_{m_{2}\to0}\frac{\theta}{2\pi}=\frac{1}{2}w_{3D}\mathrm{sgn}(m_{2}).
\end{equation}
Hence, $\frac{\theta}{2\pi}$ is quarter quantized and is one half
of the winding number with an additional factor $\mathrm{sgn}(m_{2})$
as $m_{2}\to0$.

Besides, we can also calculate the magnetoelectric effect in a finite
magnetic field, where Landau levels are formed. To find the solution
of Landau levels, we make a unitary transformation $\mathcal{U}$
for the original Hamiltonian as $\mathcal{U}^{\dagger}H\mathcal{U}=-\tau_{2}b(\Pi\cdot\sigma^{\prime})^{2}+\tau_{1}v\Pi\cdot\sigma^{\prime}-m_{2}\tau_{3}$.
The momentum operators $\hbar k_{i}$ are replaced by kinematic momentum
operators $\Pi_{i}=\hbar k_{i}+eA_{i}$ under the Pierls substitution,
where $A_{i}$ is $i$th component of the vector potential. Without
loss of generality, one can choose the vector potential as $A_{i}=-\delta_{i1}By$
and $\delta_{ij}$ the Kronecker delta symbol. By taking advantage
of the ladder operator technique \citep{shen2005prb}, we can obtain
the eigenvalues and eigenstates for the operator, 
\begin{equation}
\Pi\cdot\sigma^{\prime}\left|nk_{x}k_{z}\chi_{n}\right\rangle =\chi_{n}\sqrt{\hbar^{2}k_{z}^{2}+n\hbar^{2}\Omega^{2}/v^{2}}\left|nk_{x}k_{z}\chi_{n}\right\rangle 
\end{equation}
where $n=0,1,2,\cdots$ are the indices of the Landau levels, and
$\Omega\equiv\sqrt{2}v\ell_{B}^{-1}$ is the cyclotron frequency with
the magnetic length $\ell_{B}=\sqrt{\hbar/eB}$ . $\chi_{n}$ stands
for the helicity of massive Dirac fermions: $\chi_{n}=\pm1$ for $n>0$
and $\chi_{0}=\mathrm{sgn}(k_{z})$ for $n=0$. Then, the energy spectra
are quantized into a series of Landau levels. In fact, the lowest
Landau levels are primarily responsible for the emergence of exotic
quantum phenomena. In this way, we can reduce the dimension to one
by projecting onto the lowest landau levels with the Landau degeneracy
$eB/\hbar$. The Hamiltonian and all the physical quantities can be
expressed in terms of 2 \texttimes{} 2 matrices and the motions of
electrons are confined along the direction of magnetic field, \textit{i.e.},
\begin{equation}
H^{(0)}=v\hbar k_{z}\tau_{1}-b\hbar^{2}k_{z}^{2}\tau_{2}-m_{2}\tau_{3}.\label{eq:1d_Hamiltonian}
\end{equation}
The eigen states of the one-dimensional Hamiltonian are given by $[se^{-i\varphi}\cos\frac{\phi_{s}}{2},\sin\frac{\phi_{s}}{2}]^{T}$
for $m_{2}<0$ and $[s\cos\frac{\phi_{s}}{2},e^{i\varphi}\sin\frac{\phi_{s}}{2}]^{T}$
for $m_{2}>0$, where $e^{i\varphi}\equiv\frac{\mathrm{sgn}(k_{z})[v-ib\hbar k_{z}]}{\sqrt{v^{2}+\left(b\hbar k_{z}\right)^{2}}}$,
$\cos\phi_{s}=-\frac{sm_{2}}{\sqrt{\left(v\hbar k_{z}\right)^{2}+\left(b\hbar^{2}k_{z}^{2}\right)^{2}+m_{2}^{2}}}$,
$s=+$ for conduction band and $s=-$ for valence band. For a given
$m_{2},$ the eigen state is always well-defined in the momentum space,
and the corresponding electric polarization $-e\left\langle i\partial_{3}\right\rangle $
can be calculated as $P_{z}^{(0)}=-e\mathrm{sgn}(m_{2})\int_{-\infty}^{+\infty}\partial_{k_{z}}\varphi\cos^{2}\frac{\phi_{\mathrm{sgn}(m_{2})}}{2}\frac{dk_{z}}{2\pi}$,
which is quarter quantized as $P_{z}^{(0)}=\frac{e}{4}\mathrm{sgn}(m_{2}b)$
as $m_{2}\to0$. Consider the Landau degeneracy $n_{B}=|eB|/2\pi\hbar$,
the variation of the free energy density from the lowest Landau levels
can be found as $\Delta\mathcal{F}=eP_{z}^{(0)}\mathbf{E}\cdot\mathbf{B}/h$.
For higher Landau levels, the contribution can be proved to vanish
in the limit of $m_{2}\to0$. Consequently, the total $\Delta\mathcal{F}$
becomes $\Delta\mathcal{F}=e^{2}\mathrm{sgn}(m_{2}b)\mathbf{E}\cdot\mathbf{B}/4h$
as $m_{2}\to0$, which corresponds to a fractional $\theta$-term
in units of $2\pi$ as $\lim_{m_{2}\to0}\frac{\theta}{2\pi}=\frac{1}{2}w_{3D}\mathrm{sgn}(m_{2})$.
which is consistent from the results in Eq. (\ref{eq:theta_m2to0}).
In the absence of $m_{2},$ the model has the parity symmetry $\tau_{2}H^{(0)}(-k_{z})\tau_{2}=H^{(0)}(k_{z})$.
Due to the double degeneracy of the zeroth Landau bands, the two orthogonal
states at the $k_{z}=0$ can be expressed as superpositions of the
odd-parity and even-parity states. The coefficient $\mathrm{sgn}(m_{2})$
indicates that the spontaneous symmetry breaking of the electric polarization
induced by the external perturbation $m_{2}\tau_{3}\sigma_{2}$. The
sign of $m_{2}$ determines the occupancy of the degenerated states.
Thus the $\theta$ field emerges in the gapless CAS as a consequence
of spontaneous symmetry breaking induced by an infinitesimal small
field $m_{2}$.

\section{Surface Hall conductance}

The nonzero value of the field $\theta$ in the bulk means the emergence
of the surface Hall effect on the surface of the system. To explore
the surface Hall effect of CAS, the continuum Hamiltonian (\ref{eq:model})
is discretized on a cubic lattice by the replacement $k_{i}\to\sin k_{i}$
and $k_{i}^{2}\to4\sin^{2}\frac{k_{i}}{2}$ in a unit lattice spacing
as
\begin{align*}
H= & v\hbar(\sin k_{x}\tau_{3}\sigma_{3}+\sin k_{y}\tau_{2}+\sin k_{z}\tau_{3}\sigma_{1})\\
 & -4b\hbar^{2}\sum_{i=x,y,z}\sin^{2}\frac{k_{i}}{2}\tau_{1}+m_{2}\tau_{3}\sigma_{2}.
\end{align*}
The winding number for this lattice model can be numerically found
as $w_{3D}=-\frac{1}{2}\mathrm{sgn}(b)$, which is consistent with
the continuum model. Consider a slab geometry that is finite in the
$z-$direction, and perform the periodic condition along the $x$
and $y$ directions. The surface Hall effect can be resolved by the
layer-dependent Hall conductance as \citep{Essin2009OMP,wang2015magnetoelectric}
\begin{equation}
\sigma_{xy}(\ell)=\frac{e^{2}}{h}\frac{1}{2\pi}\sum_{\ell^{\prime}}\int dk_{x}dk_{y}\sum_{m,n}\frac{2\mathrm{Im}[v_{nm}^{x}(\ell)v_{mn}^{y}(\ell^{\prime})]}{(E_{n}-E_{m})^{2}},
\end{equation}
where $\ell$ labels the layer index along the $z$ direction, $n$
and $m$ label the occupied and unoccupied bands, respectively. $v_{nm}^{i}(\ell)\equiv\langle\varphi_{n}(k_{x},k_{y})|\frac{1}{\hbar}\frac{\partial H_{\ell}}{\partial k_{i}}|\varphi_{m}(k_{x},k_{y})\rangle$
are the matrix elements of local velocity operators along the $i-$direction
with $i=x,y$, $H_{\ell}=\sum_{s,s^{\prime}}\langle\ell,s|(H_{0}+m_{2}\tau_{3}\sigma_{2})|\ell,s^{\prime}\rangle|\ell,s\rangle\langle\ell,s^{\prime}|$
projects out the $\ell$th sector of the Hamiltonian , $s$ and $s^{\prime}$
denote the spin or orbital indices. According to the topological field
theory \citep{Qi2008QFT}, the value of $\theta$ in the vacuum is
$\theta=0$. The axion field will decay near the sample surface, and
the bulk properties of $\theta$ in unit of $2\pi$ can be deducted
from the surface Hall conductance $\sigma_{xy}$ 
\begin{equation}
\frac{\theta_{f}}{2\pi}=-\frac{h}{e^{2}}\sum_{\ell^{\prime}=1}^{N_{L}/2}\sigma_{xy}(\ell^{\prime}).\label{eq: theta}
\end{equation}
Here we have used the symbol $\theta_{f}$ to distinguish with the
$\theta$ calculated from the periodic boundary condition, and the
subscript $f$ means finite size. $N_{L}$ is the total number of
layers along the $z$-direction. For a slab with finite thickness,
$\theta_{f}$ is a function of $N_{L}$. In the thermodynamic limit
$N_{L}\to+\infty$, $\theta_{f}$ becomes $\theta$ \citep{wang2015magnetoelectric}.
It is noted that $\frac{\theta}{2\pi}$ evaluated from the non-Abelian
Berry curvature is ambiguous up to an integer, while $\frac{\theta_{f}}{2\pi}$
from Eq. (\ref{eq: theta}) is always unambiguous. Hence, it is more
proper to define the $\theta$-term from the layer-dependent Hall
conductance in real materials.

\begin{figure}
\centering{}\includegraphics[width=8.5cm]{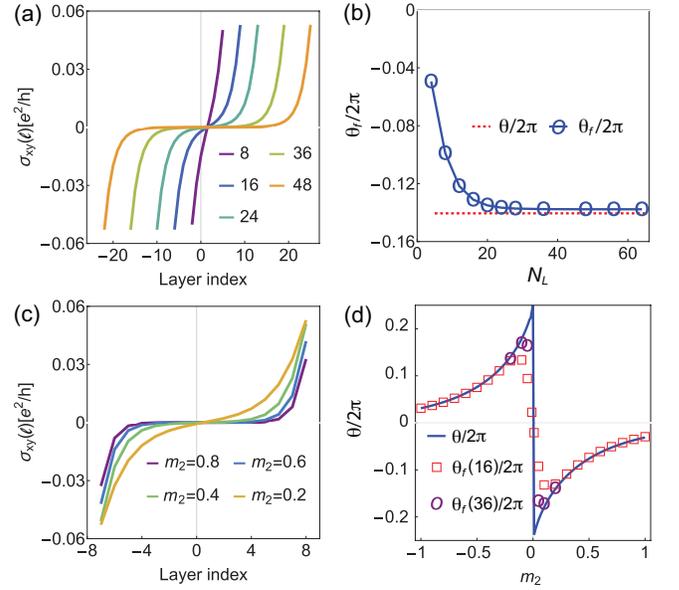}\caption{\label{fig:(a)Layer-dependent-Chern-number}(a) Layer-dependent Hall
conductance in units of $e^{2}/h$ as a function of layer index for
CAS with different total layer numbers ($N_{L}=8,16,24,36,48$). (b)
The extracted $\theta_{f}$ (blue circles) as function of $N_{L}$
and $\theta$ (red dashed line). (c) Layer-dependent Hall conductance
for CAS with different $m_{2}$ ($0.2,0.4,0.6,0.8$). (d) Comparison
of $\theta$ and extracted $\theta_{f}$ as functions of $m_{2}$.
In (a) and (b), $m_{2}$ is fixed as $m_{2}=0.2$. In (c), $N_{L}$
is fixed as $N_{L}=16$. Other calculation parameters are chosen as
$b\hbar^{2}=0.5$ and $v\hbar=1$ in (a-d).}
\end{figure}

To reveal the relation between $\theta$ and $\theta_{f},$ we calculate
the layer-dependent Hall conductance for CAS state with finite $m_{2}$.
As shown in Fig. \ref{fig:(a)Layer-dependent-Chern-number}(a), we
has evaluated the layer-dependent Hall conductance $\sigma_{xy}(\ell)$
as a function of layer index $\ell$ for several different total layer
numbers $N_{L}=8,16,24,36,48$, and the layer index is labeled from
$-N_{L}/2+1$ to $N_{L}/2$. It is noted that the $\sigma_{xy}(\ell)$
is localized around the surface. Its magnitude has a maximum at the
top and bottom layers and gradually decreases to $0$ in the bulk,
and the corresponding penetration depth is around 10 layers for the
selected model parameters. As depicted in Fig. \ref{fig:(a)Layer-dependent-Chern-number}(b),
the dimensionless parameters $\theta_{f}$ deduced from Fig. \ref{fig:(a)Layer-dependent-Chern-number}(a)
tends to be saturated to a constant $\theta$ by fixing $m_{2}=0.2$
when the total layer number $N_{L}\to+\infty$. For a given total
layer number, for example, $N_{L}=16$, the penetration depth of layer
Hall conductance decreases with increasing $m_{2}$ as shown in Fig.
\ref{fig:(a)Layer-dependent-Chern-number}(c). Meanwhile, the extract
$\theta_{f}$ from the $\sigma_{xy}(\ell)$ is more closed to $\theta$
with increasing $m_{2}$ as shown in Fig. \ref{fig:(a)Layer-dependent-Chern-number}(d).
For small $m_{2},$ a larger layer number is required to make $\theta_{f}$
approaching $\theta$.

\begin{figure}
\centering{}\includegraphics[width=8.5cm]{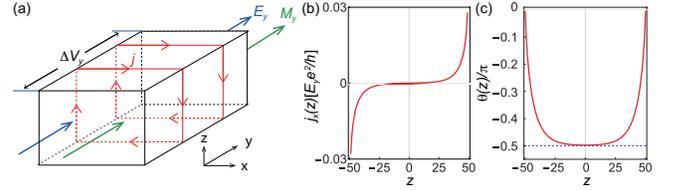}\caption{\label{fig:current_distribution}(a)Schematic diagram of the relation
between magnetization and bulk topological magnetoelectric effect.
A surface current is produced by an electric field due to the magnetization.
(b)Local current density along the $x-$direction as a function of
slab position $z$. (c) Spatial dependent $\theta$ along the z direction.
The electric field is applied along the $y$-direction. The calculation
parameters are chosen as $v\hbar=1$, $m_{2}=0.05$ and $L_{z}=101$.}
\end{figure}

Opposite the conventional topological insulator, there is no corresponding
well-defined surface states for CAS with finite $m_{2}$, although
there is a layer-dependent Hall conductance. The $m_{2}$ term breaks
the time reversal symmetry and causes a bulk band gap. The generation
of electric current near the sample surface can be regarded as a consequence
of the magnetoelectric effect. For illustration, we consider a cuboid
geometry in Fig. \ref{fig:current_distribution}(a), and there is
a voltage difference $\Delta V_{y}$ along the $y-$direction, which
corresponds to a constant electric field $E_{y}$ (blue arrow). Analytically,
the current distribution and orbital magnetization can be calculated
by employing the standard perturbation theory. Here we consider the
case of $m_{2}\to0$, for a slab with thickness $L_{z}$, we obtain
the spatial dependence of current density as (see Appendix \ref{sec:The-topological-magnetoelectric}
for details), 
\begin{equation}
j_{x}(z)=\frac{e^{2}E_{y}}{2\pi\hbar}\frac{m_{2}}{\pi\hbar v}\sum_{\pm}\pm K_{0}[|\frac{2m_{2}}{\hbar v}(\frac{L_{z}}{2}\pm z)|],\label{eq:current_density}
\end{equation}
where $K_{0}[x]$ is the zeroth order modified Bessel function of
the second kind with argument $x$. Position $z$ varies from $-L_{z}/2$
to $L_{z}/2$. It is easy to see that the value of current is asymmetric
about $z=0$, namely, $j_{x}(z)=-j_{x}(-z)$, and the total current
density is zero for the whole bar sample, which is consistent with
the layer Hall conductance in Fig. \ref{fig:(a)Layer-dependent-Chern-number}(a).
As shown in Fig. \ref{fig:current_distribution}(b), the local current
density mainly distributes near the top and bottom few layers, and
nearly vanishes inside the deep bulk. Taking advantage of the asymptotic
form of $K_{0}(x)$ at a small argument $K_{0}(x)\sim-\ln x$, we
find the current $j_{x}^{\pm}(z)\sim\pm\frac{e^{2}E_{y}}{2\pi\hbar}\frac{m_{2}}{\pi\hbar v}\ln|\frac{2m_{2}}{\hbar v}(\frac{L_{z}}{2}\pm z)|$,
which is logarithmic decay near the slab surface. Such a logarithmic
decay behavior is different from the exponential decay one in topological
insulators and is a unique property of CAS with $m_{2}$. When $z$
deviates from the surface and $|2m_{2}(\frac{L_{z}}{2}\pm z)|$ becomes
large, $j_{x}(z)$ develops into exponential decay behavior along
the $z$ direction as $j_{x}^{\pm}(z)\sim\mp\frac{e^{2}}{2\pi\hbar}\frac{m_{2}}{2\sqrt{\pi\hbar v}}\frac{e^{-|2\frac{m_{2}}{\hbar v}(z\pm\frac{L_{z}}{2})|}}{\sqrt{|m_{2}(z\pm\frac{L_{z}}{2})|}}$,
and the decaying length is proportional to the inverse of the energy
gap $m_{2}$. It should be emphasized that the surface current is
totally contributed from the extended bulk states as there is no localized
surface states in CAS. The external electric field mixes the electrons
and holes to generate a finite current at the half-filling even there
is no states within the bandgap. A similar case has been studied in
the two dimensional parity anomalous semimetal, where the accumulation
of extended bulk states exhibit chiral nature and an edge current
flows along the massive/massless domain wall \citep{Zou2022arxiv}.

Furthermore, the axion action $S_{\theta}=\frac{e^{2}}{h}\int d^{3}xdt\frac{\theta}{2\pi}\mathbf{E}\cdot\mathbf{B}$
tells us that the electric field will produce a magnetization as $\mathbf{M}=\theta e^{2}/(4\pi^{2}\hbar)\mathbf{E}$
(green arrow); then, the magnetization will produce a current near
the surface as $\mathbf{j}=\nabla\times\mathbf{M}=e^{2}/(4\pi^{2}\hbar)\nabla\theta\times\mathbf{E}$
(red arrow), which is perpendicular to the direction of electric field
\citep{Qi2008QFT}. Inversely, the current density (Eq. (\ref{eq:current_density}))
corresponds to a spatially varied orbital magnetization, $M_{y}(z)=-\int_{-L_{z}/2}^{z}dz^{\prime}j_{x}(z^{\prime})$.
Then, we can obtain the spatial dependent $\theta$ along the $z$
direction as 
\begin{equation}
\frac{\theta(z)}{2\pi}=\frac{\mathrm{sgn}(m_{2})}{4}\left(\mathcal{F}(|\frac{2m_{2}L_{z}}{\hbar v}|)-\sum_{s=\pm}\mathcal{F}(|\frac{m_{2}}{\hbar v}(2sz+L_{z}|))\right)\label{eq:theta_z}
\end{equation}
where $\mathcal{F}(x)=x(K_{0}(x)H_{-1}(x)+K_{1}(x)H_{0}(x))$ with
$H_{n}(x)$ the Struve functions with argument $x$. $\mathcal{F}(x)$
has the asymptotic behavior $\mathcal{F}(x)\sim1$ at large argument.
Hence, in the thermodynamic limit $2m_{2}L_{z}/\hbar v\to\infty$,
the bulk value $\theta(z=0)$ converges to $\pi/2$ as shown in Fig.
\ref{fig:current_distribution}(c).

\begin{figure}
\begin{centering}
\includegraphics[width=8.5cm]{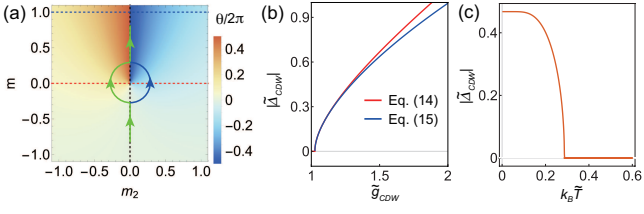}
\par\end{centering}
\caption{\label{fig:Phase-diagram-of}(a) Phase diagram of $\theta$ as functions
of $m_{2}$ and $m$. Other parameters are chosen as $v\hbar=1$ and
$b\hbar^{2}=0.5$. (b)The amplitude of $\tilde{\Delta}_{SDW}$ at
zero temperature as a function of dimensionless coupling constant
$\tilde{g}_{SDW}$. The red line is solved numerically from Eq. (\ref{eq:zerotemperature}),
the blue line is calculated in terms of the power-law function in
Eq. (\ref{eq:order_parameter}), where the dimensionless mass are
chosen as $\tilde{m}=-0.2$. (c) The amplitude of $\tilde{\Delta}_{SDW}$
as a function of dimensionless temperature $k_{B}\tilde{T}$, which
is solved numerically from Eq. (\ref{eq:finite_temperature-1}). The
dimensionless mass and coupling constant are chosen as $\tilde{m}=-0.2$
and $\tilde{g}_{SDW}=1.3$, respectively.}
\end{figure}

For comparison, we also present the results for topological insulators
by considering $m\ne0$ in Eq. (\ref{eq:model}). It is known that
$mb>0$ describes the topological insulators and $mb<0$ describes
the trivial insulators \citep{lu2010prb}. For nonzero $m_{2}$, $\frac{\theta}{2\pi}$
is a function of m and $m_{2}$ for a given $b$ as shown in Fig.
\ref{fig:Phase-diagram-of}(a), and it always satisfies the relation
as $\frac{\theta}{2\pi}\propto\mathrm{sgn}(m_{2})$. When $m_{2}\to0$,
$\frac{\theta}{2\pi}$ can be analytically found as $\frac{\theta}{2\pi}=-\frac{1}{4}[\mathrm{sgn}(m)+\mathrm{sgn}(b)]\mathrm{sgn}(m_{2})$.
Hence, as $m_{2}\to0$, $\frac{\theta}{2\pi}$ is half-quantized as
$\pm1/2$ or $0$ for a finite $m$ and $b$, while $\frac{\theta}{2\pi}$
is quarter-quantized as $\pm\frac{1}{4}$ for $m=0$.

Physically, the action $S_{\theta}$ modifies the Maxwell's equations
and leads to unusual optical properties, such as the topological Kerr
and Faraday rotations, which provide an effective way to measure the
magnetoelectric polarization in solids \citep{maciejko2010prl,Tse2011prb,fu2021prr}.
Near the boundary between CAS ($\theta_{CAS}$) and vacuum ($\theta=0$),
the surface Hall conductance becomes $\sigma_{xy}^{T(B)}=\frac{e^{2}}{h}\nu_{T(B)}$,
where $\nu_{T}=-\frac{\theta_{CAS}}{2\pi}$ and $\nu_{B}=\frac{\theta_{CAS}}{2\pi}$.
The two surface Hall conductance can be related to the magneto-optical
measurement. One can extract the information of $\theta_{CAS}$ through
an optical measurement. In experiments, there is a normally incident
linearly $x$-polarized light propagating along the $z$ direction.
The Kerr and Faraday angle at reflectivity minimum and maximum are
obtained as $\theta_{K,F}^{\prime}$ and $\theta_{K,F}^{\prime\prime}$,
respectively. As $\nu_{B}+\nu_{T}=0$, we have $2\alpha P_{3}^{CAS}=\tan(\theta_{K}^{\prime\prime}+\theta_{F}^{\prime\prime})$
\citep{fu2021prr}. We expect the obtained $\theta_{CAS}$ approaches
the quantized value $\pm\pi/2$ as $m_{2}\to0$, which can provide
a substantial evidence for the existence of CAS in solids.

\section{Physical realization of $m_{2}$}

The symmetry breaking term $m_{2}\tau_{3}\sigma_{2}$ can be realized
by a staggered Zeeman field pointing in the $y$ direction on the
sublattice $\mathrm{Te_{1}}$ and $\mathrm{Te_{2}}$ of each unit
cell in transition-metal pentatelluride $\mathrm{ZrTe}_{5}$. This
Zeeman field can be produced by a spin density wave order, where spins
point along opposite $y$ directions on two sublattices. In the mean
field approximation, the system can develop spin density wave order
$\Delta_{SDW}$ if the effect of on-site repulsion $H_{U}=U\sum_{i}(n_{iA\uparrow}n_{iA\downarrow}+n_{iB\uparrow}n_{iB\downarrow})$
dominates that of inter-site repulsion $H_{V}=V\sum_{i,\sigma}n_{iA\sigma}n_{iB\sigma}$
with $A/B=\mathrm{Te_{1/2}}p_{y}$ \citep{Li2016np,sekine2014jpsj}.
As the spin density wave order enters the Hamiltonian (\ref{eq:model})
as $\delta H=\Delta_{SDW}\tau_{3}\sigma_{2}$, it indeed produces
the desired symmetry breaking term \citep{Aoki1986prl}. Moreover,
we will show that such a spin density wave order can be tuned by external
strain. By introducing the dimensionless parameters $\widetilde{g}_{SDW}=\frac{U\Lambda^{2}}{16\pi^{2}(\hbar v)^{3}}$,
$\widetilde{\Delta}_{SDW}=\frac{\Delta_{SDW}}{\Lambda}$ , $\widetilde{m}=m/\Lambda$,
$\widetilde{T}=T/\Lambda$ with $\Lambda$ the energy cutoff, the
self-consistent equation for $\tilde{\Delta}_{SDW}$ becomes (see
Appendix \ref{sec:Spin-density-wave} for details)
\begin{align}
\widetilde{\Delta}_{SDW} & =2\widetilde{g}_{SDW}\int_{0}^{1}dxx^{2}\frac{\widetilde{\Delta}_{SDW}}{\sqrt{x^{2}+\widetilde{m}^{2}+\widetilde{\Delta}_{SDW}^{2}}}\label{eq:finite_temperature-1}\\
 & \times\tanh\frac{\sqrt{x^{2}+\widetilde{m}^{2}+\widetilde{\Delta}_{SDW}^{2}}}{2k_{B}\widetilde{T}},\nonumber 
\end{align}
where $k_{B}$ is the Boltzmann constant. At zero temperature, $\tanh\frac{\sqrt{x^{2}+\widetilde{m}^{2}+\widetilde{\Delta}_{SDW}^{2}}}{2k_{B}\widetilde{T}}=1$,
the above integral can be done analytically as
\begin{align}
\widetilde{\Delta}_{SDW}= & \widetilde{g}_{SDW}\widetilde{\Delta}_{SDW}(\sqrt{\widetilde{\Delta}_{SDW}^{2}+\widetilde{m}^{2}+1}\label{eq:zerotemperature}\\
 & -(\widetilde{\Delta}_{SDW}^{2}+\widetilde{m}^{2})\tanh^{-1}\frac{1}{\sqrt{\widetilde{\Delta}_{SDW}^{2}+\widetilde{m}^{2}+1}}).\nonumber 
\end{align}
The critical value for the coupling strength is determined as 
\[
\widetilde{g}_{SDW}^{c}(\widetilde{m})=(\sqrt{\widetilde{m}^{2}+1}-\widetilde{m}^{2}\tanh^{-1}\frac{1}{\sqrt{\widetilde{m}^{2}+1}})^{-1},
\]
which is a monotonic increasing function of $|\widetilde{m}|$. Depending
on the coupling strength $\widetilde{g}_{SDW}$, there exists two
distinct regimes as we change the strain (which modifies the Dirac
mass effectively). When $\tilde{g}_{SDW}<1$, as $\widetilde{g}_{SDW}^{c}(\widetilde{m})\ge\widetilde{g}_{SDW}^{c}(0)=1,$
$\widetilde{g}_{SDW}^{c}(\widetilde{m})>\widetilde{g}_{SDW}$ is satisfied
regardless of the mass $\tilde{m}$, $\widetilde{\Delta}_{SDW}$ is
always zero by varying the strain. As shown in Fig. \ref{fig:Phase-diagram-of}(b),
when $\widetilde{g}_{SDW}>1$, there exists the spin density wave
instability once $\widetilde{g}_{SDW}>\widetilde{g}_{SDW}^{c}$. Considering
the order parameter and Dirac mass are small compared with $\Lambda$,
the $\tilde{\Delta}_{SDW}$ at zero temperature can be solved approximately
as
\begin{equation}
\widetilde{\Delta}_{SDW}^{2}+\widetilde{m}^{2}\approx(\widetilde{g}_{SDW}-1)^{1.266}.\label{eq:order_parameter}
\end{equation}
For comparison, we plot the numerical (red line) and analytical (blue
line) results for the amplitude of $\widetilde{\Delta}_{SDW}$ as
a function of $\tilde{g}_{SDW}$ in Fig. \ref{fig:Phase-diagram-of}(b).
Near $\tilde{g}_{SDW}^{c}$, Eq. (\ref{eq:order_parameter}) provides
a good approximation for the amplitude of order parameters.

Without loss of generality, we can assume the system is in a phase
weak topological insulator and $\widetilde{g}_{SDW}^{c}(\widetilde{m})>\widetilde{g}_{SDW}$
in the absence of strain. By increasing the strain, the $|\widetilde{m}|$
and $\widetilde{g}_{SDW}^{c}(\widetilde{m})$ decrease simultaneously
when $\tilde{m}<0$. When $|\tilde{m}|$ is smaller than the critical
value $|\widetilde{m}_{c}|$ with $\widetilde{g}_{SDW}^{c}(\widetilde{m}_{c})=\widetilde{g}_{SDW}$,
the order parameter $\widetilde{\Delta}_{SDW}$ becomes nonzero and
breaks the time-reversal and inversion symmetry spontaneously. With
further increasing of the strain, $\widetilde{\Delta}_{SDW}$ grows
following the relation (\ref{eq:order_parameter}). Then, the corresponding
axion field will be tuned along the green or blue line in Fig. \ref{fig:Phase-diagram-of}(a),
which is a semicircle as $|\widetilde{m}|<|\widetilde{m}_{c}|$. For
vanishing Dirac mass $\widetilde{m}=0$, $\widetilde{\Delta}_{SDW}$
reaches its maximum. When the strain is large enough that the $\widetilde{m}\ge|\widetilde{m}_{c}|$,
$\widetilde{\Delta}_{SDW}$ vanishes again and the system becomes
a strong topological insulator. It is noted that the bulk energy gap
closing is avoided due to the finite spin density wave order in the
process of topological phase transition.

Furthermore, the temperature dependence of the radius in Eq. (\ref{eq:order_parameter})
can be obtained numerically from finite temperature self-consistent
equation (\ref{eq:finite_temperature-1}) as shown in Fig. \ref{fig:Phase-diagram-of}(c).
The order parameter decreases with increasing the temperature below
the critical temperature, which might provide another way to tune
the axion field in Fig. \ref{fig:Phase-diagram-of}(a) by reducing
the radius of green and blue line. Hence, the CAS states near $\tilde{m}=0$
has a natural advantage in generating axion field by introducing interaction,
strain and temperature.

\section{Summary}

In this work, we studied the topological properties for the CAS. Firstly,
the effective model Hamiltonian of transition-metal pentatelluride
$\mathrm{ZrTe}_{5}$ has been presented, and the mass of $\mathrm{ZrTe}_{5}$
can be tuned to zero by strain to form CAS. Secondly, the continuity
equation {[}Eq. (\ref{eq:chiral anomaly}){]} of chiral current has
been derived. This equation depends on the fermi wavevector as $1/\sqrt{1+(b\hbar k_{F}/v)^{2}}$
and gives the continuity equation of chiral anomaly in the limit of
$k_{F}\to0$ or $b\to0$, where the chiral symmetry is restored. Besides,
the winding number for CAS has been found to be half-quantized as
$w_{3D}=-\mathrm{sgn}(b)/2$, which is distinct to the existing topological
states. In the presence of symmetry breaking term $m_{2}\tau_{3}\sigma_{2}$,
the magnetoelectric effect is found as a function of $m_{2}$. We
have analytically proved that the $\theta$ field is quarter-quantized
as $\frac{\theta}{2\pi}=\frac{1}{2}w_{3D}\mathrm{sgn}(m_{2})$ in
the limit of $m_{2}\to0$, which is totally different from the half-quantized
$\theta$ in topological insulators. Furthermore, we have calculated
the surface Hall conductance on a lattice, which has a good agreement
with the bulk value of $\theta$ for a large layer number. Opposite
to the topological insulator, there is no well-defined surface states
for CAS with finite $m_{2}$, where the nonzero surface Hall conductance
or surface current is from the collective effect of extended bulk
states not the localized surface states. Lastly, the spin density
wave order $\Delta_{SDW}$ provides a physical realization for the
$m_{2}$ term near the phase transition point in $\mathrm{ZrTe_{5}}$,
and the magnitude of $m_{2}$ can be tuned by interaction, strain
and temperature.

\section*{Acknowledgments}

This work was supported by the National Key R\&D Program of China
under Grant No. 2019YFA0308603 and the Research Grants Council, University
Grants Committee, Hong Kong under Grant No. C7012-21G and No. 17301220.

H.W.W and B.F. contributed equally to this work.

\appendix

\section{Continuity equation for chiral current\label{sec:Continuity-equation-for}}

The chiral anomaly of three dimensional Dirac fermion in a quantized
field can be reduced to the lowest Landau level, which is non-degenerate.
To understand the quantum anomaly in these systems, it is convenient
to begin with the one-dimensional case $H^{(0)}=v\hbar k_{z}\gamma^{0}\gamma^{3}-b\hbar^{2}k_{z}^{2}\gamma^{0}$
with $\gamma^{0}=\tau_{2},$ $\gamma^{0}\gamma^{3}=\tau_{1}$. We
focus on the chiral operator $\gamma^{5}=\tau_{1}$ in this section.
Following the Jackiw-Johnson approach to the chiral anomaly, we can
derive the continuity equation for the gauge-invariant chiral current
$j_{5}^{\mu}(x,\epsilon)$ by taking the limit $\epsilon\to0$ as
\begin{equation}
\partial_{\mu}j_{5}^{\mu}=-\lim_{\epsilon\to0}i\frac{e}{\hbar}\epsilon^{\alpha}F_{\alpha\mu}j_{5}^{\mu}(x,\epsilon)-i\bar{\psi}2b\hbar k_{z}^{2}\gamma^{5}\psi,\label{eq:chiral_current}
\end{equation}
where $F_{\alpha\mu}=\partial_{\alpha}A_{\mu}-\partial_{\mu}A_{\alpha}$
is the field strength, $A_{\mu}=(\phi,0)$ with $\phi$ the electric
potential. $\psi$ and $\bar{\psi}=\psi^{\dagger}\gamma^{0}$ are
the Dirac spinors. The first term gives the anomalous correction due
to the spontaneous chiral symmetry breaking. The second term is the
pseudo scalar condensation, and is the consequence of the explicit
chiral symmetry breaking from the Dirac mass $-b\hbar^{2}k_{z}^{2}$.
The gauge-invariant chiral currents are defined as
\begin{align}
j_{5}^{0}(x,\epsilon) & =e^{i\frac{e}{\hbar}\int_{x-\epsilon/2}^{x+\epsilon/2}A_{\alpha}(x)dx^{\alpha}}\bar{\psi}_{+}\gamma^{0}\gamma^{5}\psi_{-}\label{eq:chiral density}\\
j_{5}^{3}(x,\epsilon) & =e^{i\frac{e}{\hbar}\int_{x-\epsilon/2}^{x+\epsilon/2}A_{\alpha}(x)dx^{\alpha}}\{v\bar{\psi}_{+}\gamma^{3}\gamma^{5}\psi_{-}\label{eq:chial current}\\
 & +i\hbar b[\bar{\psi}_{+}\gamma^{5}(\partial_{z}\psi_{-})-(\partial_{z}\bar{\psi}_{+})\gamma^{5}\psi_{-}]\}\nonumber 
\end{align}
where $\psi_{\pm}=\psi(x\pm\frac{\epsilon}{2})$, $\bar{\psi}_{\pm}=\bar{\psi}(x\pm\frac{\epsilon}{2})$.
The exponential factor makes the current operator to be locally gauge
invariant under the transformation $A_{\mu}^{\prime}=A_{\mu}+\partial_{\mu}\chi$
and $\psi^{\prime}=\psi e^{ie\chi/\hbar}.$

Here we consider a nonzero $\epsilon_{3}$, the possible nonvanishing
contribution is from $\langle\epsilon^{3}F_{30}j_{5}^{0}\rangle$.
When $b\ne0$, 
\begin{align*}
 & \lim_{\epsilon\to0}i\frac{e}{\hbar}\epsilon^{3}F_{30}\langle j_{5}^{0}(x,\epsilon)\rangle\\
= & 2\frac{eE_{z}}{\hbar}\lim_{\epsilon_{3}\to0}\int_{0}^{+\infty}\frac{dk_{z}}{2\pi}\frac{\epsilon_{3}\sin\epsilon_{3}k_{z}}{\sqrt{1+(b\hbar k_{z}/v)^{2}}}\\
= & 2\frac{eE_{z}}{\hbar}\lim_{\epsilon_{3}\to0}\int_{0}^{+\infty}\frac{dx}{2\pi}\frac{\sin x}{\sqrt{1+(\frac{b\hbar}{v\epsilon_{3}})^{2}x^{2}}}\\
= & \lim_{\epsilon_{3}\to0}\frac{eE_{z}}{2\hbar}|\frac{v\epsilon_{3}}{b\hbar}|\left(I_{0}\left(|\frac{v\epsilon_{3}}{b\hbar}|\right)-L_{0}\left(|\frac{v\epsilon_{3}}{b\hbar}|\right)\right)\\
= & 0
\end{align*}
where $I_{0}(x)$ and $L_{0}(x)$ are the modified Bessel function
of the first kind and modified Struve functions. In the limit $x\to0$,
$I_{0}\left(x\right)-L_{0}\left(x\right)\to1$; and $\lim_{\epsilon\to0}i\frac{e}{\hbar}\epsilon^{3}F_{30}\langle j_{5}^{0}(x,\epsilon)\rangle=0$.
Hence, there is no anomalous contribution from the spontaneous chiral
symmetry breaking in Eq. (\ref{eq:chiral_current}).
\begin{widetext}
Then, let us calculate the expectation value of $\bar{\psi}2ib\hbar k_{z}^{2}\gamma^{5}\psi$
at the zero temperature and fermi energy $\mu$. In the linear response
theory, the expectation value of a general operator $\bar{\psi}\hat{\mathcal{O}}\psi$
can be computed as

\begin{align*}
\langle\bar{\psi}\hat{\mathcal{O}}\psi\rangle= & E_{z}(\Lambda^{RA}+\Lambda^{RR}-\Lambda^{AA})
\end{align*}
with
\begin{align*}
\Lambda^{RA} & =\frac{e\hbar}{2\pi L_{z}}\frac{1}{\Omega}\sum_{k_{z}}\int_{-\infty}^{\infty}d\omega\frac{n_{F}(\omega+\Omega)-n_{F}(\omega)}{\Omega}\mathrm{Tr}[\hat{\mathcal{O}}G^{R}(k_{z},\omega+\Omega)\hat{v}_{z}G^{A}(k_{z},\omega)],\\
\Lambda^{RR} & =\frac{e\hbar}{2\pi L_{z}}\frac{1}{\Omega}\sum_{k_{z}}\int_{-\infty}^{\infty}d\omega n_{F}(\omega)\mathrm{Tr}[\hat{\mathcal{O}}G^{R}(k_{z},\omega+\Omega)\hat{v}_{z}G^{R}(k_{z},\omega)],\\
\Lambda^{AA} & =\frac{e\hbar}{2\pi L_{z}}\frac{1}{\Omega}\sum_{k_{z}}\int_{-\infty}^{\infty}d\omega n_{F}(\omega+\Omega)\mathrm{Tr}[\hat{\mathcal{O}}G^{A}(k_{z},\omega+\Omega)\hat{v}_{z}G^{A}(k_{z},\omega)].
\end{align*}
Here $\hat{v}_{z}=v\gamma^{3}-2b\hbar k_{z}$ is the velocity operator,
$G^{R/A}(k_{3},\omega)=i\frac{(\omega\pm i\delta)\gamma^{0}-v\hbar k_{z}\gamma^{3}-b\hbar^{2}k_{z}^{2}}{(\omega\pm i\delta)^{2}-(v^{2}\hbar^{2}k_{z}^{2}+b^{2}\hbar^{4}k_{z}^{4})}$
are the retarded/advanced Green's functions. $n_{F}(\omega)=\Theta(\mu-\omega)$
is the Fermi-Dirac distribution function at zero temperature. $L_{z}$
is the length of the effective one dimensional system.
\end{widetext}

For $\hat{\mathcal{O}}=-2ib\hbar k_{z}^{2}\gamma^{5}$, after a tedious
but straightforward calculation, we have 
\begin{align*}
\lim_{\Omega\to0}\Pi^{RA}= & \frac{e}{\pi\hbar}\frac{v}{v_{F}}\left(\frac{b\hbar k_{F}^{2}}{\mu}\right)^{2},\\
\lim_{\Omega\to0}(\Pi^{RR}-\Pi^{AA})= & \frac{e}{\pi\hbar}\{\frac{1}{\sqrt{1+(b\hbar k_{F}/v)^{2}}}-\frac{v}{v_{F}}\left(\frac{b\hbar k_{F}^{2}}{\mu}\right)^{2}\},
\end{align*}
with $k_{F}$ the fermi wavevector.

Then, the expectation value of pseudo scalar condensation at the dc
limit ($\Omega\to0$) becomes 
\begin{align*}
-\langle\bar{\psi}2ib\hbar k_{z}\gamma^{5}\psi\rangle= & E_{z}\lim_{\Omega\to0}(\Pi^{RA}+\Pi^{RR}-\Pi^{AA})\\
= & \frac{eE_{z}}{\pi\hbar}\frac{1}{\sqrt{1+(b\hbar k_{F}/v)^{2}}}.
\end{align*}
As only the Lowest Landau level contributes to the nonzero term in
the continuity equation, we can generalize this equation to three
dimensions by multiplying the Landau level degeneracy $n_{L}=eB/2\pi\hbar$,
which gives the continuity equation in Eq. (\ref{eq:chiral anomaly})

\section{The topological magnetoelectric effect from bulk states in CAS\label{sec:The-topological-magnetoelectric}}

We first solve the eigenenergies and eigenwavefunction without external
electric or magnetic field. Then we turn on the external field and
discuss the particle production. The appearance of surface Hall conductance
necessarily requires the breaking of time-reversal symmetry. Here,
we have introduced the $m_{2}$ term which breaks the time-reversal
and inverse symmetry simultaneously. In order to investigate the unique
bulk boundary correspondence of the chiral anomalous semimetal, we
consider a slab geometry with open boundary condition in $z$ direction
and periodic boundary condition in $x$ and $y$ direction. Thus,
the wavevectors along $x$ and $y$ directions are still good quantum
numbers. The wavefunction thus can be expressed as $|nk_{x}k_{y}\rangle=\Psi_{nk_{x}k_{y}}(z)e^{i(k_{x}x+ik_{y}y)}$
. We first solve the negative eigenvalues and the the corresponding
wavefunctions $\Psi_{nk_{x}k_{y}}$, then the positive eigenvalues
and the wavefunctions can be obtained through the particle-hole symmetry
$\Psi_{-nk_{x}k_{y}}=\Xi\Psi_{n,-k_{x},-k_{y}}$ with $\Xi$ being
the particle-hole symmetry operator. The term $b\hbar^{2}k^{2}$ behaves
as the mass regulator $\hat{M}$ and in the limit $b\to0$, it plays
an equivalent role as a hard wall boundary condition,
\begin{equation}
\left(1\pm i\hat{v}_{z}\hat{M}\right)\Psi_{nk_{x}k_{y}}(\pm L_{z}/2)=0\label{eq:BC}
\end{equation}
with $\hat{M}$ and $\hat{v}_{z}$ being the matrices for mass term
and the velocity along $z$ direction. This boundary condition breaks
the chiral symmetry explicitly. By matching the boundary conditions
in Eq. (\ref{eq:BC}), the wavefunctions can be solved as
\begin{widetext}
\begin{align*}
\Psi_{nk_{x}k_{y}}^{1} & =\frac{1}{2\sqrt{L_{z}}}\left(\left(\begin{array}{c}
-1\\
0\\
\frac{\hbar vk_{z}^{n}+im_{2}}{\varepsilon_{n}}\\
\frac{\hbar v(k_{x}+ik_{y})}{\varepsilon_{n}}
\end{array}\right)e^{ik_{z}^{n}z}-(-1)^{n}\left(\begin{array}{c}
\frac{im_{2}+\hbar vk_{z}^{n}}{\varepsilon_{n}}\\
-\frac{\hbar v(k_{x}+ik_{y})}{\varepsilon_{n}}\\
1\\
0
\end{array}\right)e^{-ik_{z}^{n}z}\right)\\
\Psi_{nk_{x}k_{y}}^{2} & =\frac{1}{2\sqrt{L_{z}}}\left(\left(\begin{array}{c}
0\\
-1\\
\frac{\hbar v(k_{x}-ik_{y})}{\varepsilon_{n}}\\
-\frac{\hbar vk_{z}^{n}-im_{2}}{\varepsilon}
\end{array}\right)e^{ik_{z}^{n}z}+(-1)^{n}\left(\begin{array}{c}
-\frac{\hbar v(k_{x}-ik_{y})}{\varepsilon_{n}}\\
-\frac{\hbar vk_{z}^{n}-im_{2}}{\varepsilon_{n}}\\
0\\
1
\end{array}\right)e^{-ik_{z}^{n}z}\right)
\end{align*}
where $k_{z}^{n}=\frac{\pi}{L_{z}}(\frac{1}{2}+n)$ and the superscripts
$1$ and $2$ denote the degenerate two states with the same eigenenergies
\[
\varepsilon_{nk_{x}k_{y}}=\sqrt{\hbar^{2}v^{2}[k_{x}^{2}+k_{y}^{2}+(k_{z}^{n})^{2}]+m_{2}^{2}}.
\]
\end{widetext}

In the presence of the electric field, we can evaluate the current
distribution and orbital magnetization by employing the standard perturbation
theory of quantum to calculate the wavefunction correction. We consider
the external electric field $\mathbf{E}(\mathbf{r})=E\cos(qy)\hat{y}$
with an infinitely slow spatial variation which recovers a uniform
electric field by taking the limit $q\to0$. The electric potential
thus can be expressed as 
\[
U=2\mathrm{Re}V(q)
\]
 with $V(q)=\frac{eEe^{-iqx}}{2iq}$ and $\mathrm{Re}$ denoting the
real part. The positive and negative energy modes provide distinct
solutions for the equation of motion for $U=0$, i.e., they do not
mix with each other. However, this is no longer true once we turn
on the electric field. There the positive and negative frequencies
get mixed due to the presence of the electric field, resulting in
the particle production. The electric field will not mix the states
between different branches $\langle snk_{x}k_{y}|V(\mathbf{r})|s^{\prime}n^{\prime}k_{x}^{\prime}k_{y}^{\prime}\rangle=0$
for $n\ne n^{\prime}$ and $n\ne-n^{\prime}$ . The $2\times2$ coupling
matrix between the positive and negative energy modes can be obtained
as,
\[
V_{q}^{n,-n}=\frac{ieE(-1)^{n}}{4\varepsilon_{nk_{x}k_{y}}\varepsilon_{n,k_{x}+q,k_{y}}}\left(\begin{array}{cc}
\hbar vk_{z}^{n}-im_{2} & \hbar v(k_{x}-ik_{y})\\
\hbar v(k_{x}+ik_{y}) & -\hbar vk_{z}^{n}-im_{2}
\end{array}\right)
\]
with the elements as $(V_{q}^{n,-n})_{ss^{\prime}}=\frac{eE}{2iq}\langle\Psi_{nk_{x}k_{y}}^{s}|\Psi_{-n,k_{x}+q,k_{y}}^{s^{\prime}}\rangle$.
Based on the perturbation theory, the first order correction to the
wavefunctions reads:
\begin{align*}
|\Phi_{n,k_{x},k_{y}}^{s^{\prime}}\rangle & \approx|\Psi_{n,k_{x},k_{y}}^{s^{\prime}}\rangle+\sum_{s=1,2}\sum_{Q=\pm q}\frac{|\Psi_{-n,k_{x}+Q,k_{y}}^{s}\rangle(V_{Q}^{n,-n})_{s,s^{\prime}}^{\dagger}}{\varepsilon_{n,k_{x},k_{y}}-\varepsilon_{-n,k_{x}+Q,k_{y}}}.
\end{align*}
The current along $y$ direction can be evaluated as 
\[
J_{y}(z)=e\sum_{n=0}^{\infty}\int\frac{dk_{x}dk_{y}}{(2\pi)^{2}}\sum_{s=1,2}\Phi_{n,k_{x},k_{y}}^{s\dagger}(z)\hat{v}_{y}\Phi_{n,k_{x},k_{y}}^{s}(z)
\]
with the current operator $\hat{v}_{y}=v\tau_{2}$. After performing
the integration over $k_{x}$ and $k_{y}$ , summation over all the
branches $n=0,1,2...$ and the degenerate two bands $s=1,2$, we obtain
Eq. (\ref{eq:chiral anomaly}) in the main text.

\section{Spin density wave order\label{sec:Spin-density-wave}}

The interactions in this model are given by
\[
H_{int}=\sum_{\boldsymbol{R}}\left[U_{1}\sum_{a,s}n_{\boldsymbol{R}as}n_{\boldsymbol{R}a\bar{s}}+U_{2}\sum_{a,s,s^{\prime}}n_{\boldsymbol{R}as}n_{\boldsymbol{R}\bar{a}s^{\prime}}\right]
\]
where $n_{\boldsymbol{R}as}=\psi_{\boldsymbol{R}as}^{\dagger}\psi_{\boldsymbol{R}as}$
is the number operator on site $\boldsymbol{R}$ for spin $s$ and
orbital $a$, and $U_{1}$ and $U_{2}$ are the intra- and inter-orbital
repulsion, respectively. By transforming to the momentum space, the
4-point interaction has the form $\sum_{\boldsymbol{k},\boldsymbol{k}^{\prime},\boldsymbol{q}}(\psi_{\boldsymbol{k}+\boldsymbol{q}}^{\dagger}\tau_{i}s_{j}\psi_{\boldsymbol{k}})(\psi_{\boldsymbol{k}^{\prime}-\boldsymbol{q}}^{\dagger}\tau_{i}s_{j}\psi_{\boldsymbol{k}^{\prime}})$
where two sets of Pauli matrices $\sigma_{j}$ and $\tau_{i}$ with
$i,j=0,1,2,3$ operate on the spin and orbital space respectively.
We only retain the terms with $\boldsymbol{q}=0$ which give rise
to the ordering between the two orbitals in the same unit cell without
spatial variation. Within the mean field, there are three possible
instabilities: ferromagnetic (FM), spin-density wave (SDW), and the
charge density wave (CDW). Thus we can decompose the 4-point interaction
into following channels,
\begin{align*}
H_{int}\sim & -\frac{1}{2V}\sum_{\boldsymbol{k},\boldsymbol{k}^{\prime}}\Bigg\{ g_{FM}\left(\psi_{\boldsymbol{k}}^{\dagger}\tau_{0}\boldsymbol{\sigma}\psi_{\boldsymbol{k}}\right)\cdot\left(\psi_{\boldsymbol{k}^{\prime}}^{\dagger}\tau_{0}\boldsymbol{\sigma}\psi_{\boldsymbol{k}^{\prime}}\right)\\
 & +g_{SDW}\left(\psi_{\boldsymbol{k}}^{\dagger}\tau_{3}\boldsymbol{\sigma}\psi_{\boldsymbol{k}}\right)\cdot\left(\psi_{\boldsymbol{k}^{\prime}}^{\dagger}\tau_{3}\boldsymbol{\sigma}\psi_{\boldsymbol{k}^{\prime}}\right)\\
 & +g_{CDW}\left(\psi_{\boldsymbol{k}}^{\dagger}\tau_{3}\sigma_{0}\psi_{\boldsymbol{k}}\right)\left(\psi_{\boldsymbol{k}^{\prime}}^{\dagger}\tau_{3}\sigma_{0}\psi_{\boldsymbol{k}^{\prime}}\right)\Bigg\}
\end{align*}
with channel couplings $g_{FM}=\frac{U_{1}}{8},g_{SDW}=\frac{U_{1}}{8},$
and $g_{CDW}=\frac{2U_{2}-U_{1}}{8}$. Here we only retain the channels
which give rise to the order parameters independent of momentum. This
quartic interaction can be decoupled by means of Hubbard-Stratonovich
transformation by introducing the real order parameter $\delta H=\sum_{\eta}\Delta_{\eta}\Gamma_{\eta}$,
where $\Delta_{\eta}$ and $\Gamma_{\eta}$ with $\eta=\mathrm{FM},\text{SDW}$
and $\text{CDW}$ are the order parameters and the corresponding $4\times4$
matrices, respectively. The mean field energies can be calculated
by the standard Ginsburg-Landau theory. By comparing the condensation
energy for different phases, we can obtain the energetically favorable
state. (i) Normal state: for relatively weak interactions, the normal
phase remains stable. These phase retains all the symmetries of the
original Hamiltonian. (ii) SDW state: in the limit of the large intra-orbital
repulsion, the SDW state with opposite spin on two orbital is stabilized.
(iii) CDW state: in the limit of the large inter-orbital repulsion,
the CDW state is expected.After performing the mean-field approximation
to the interaction term, the mean-field Hamiltonian of the system
is $\hat{\mathcal{H}}_{\mathrm{MF}}=\sum_{\boldsymbol{k}}\psi_{\boldsymbol{k}}^{\dagger}[H_{0}(\boldsymbol{k})+\delta H]\psi_{\boldsymbol{k}}$.

At half filling ($\mu=0$), the energy eigenvalues of the mean field
Hamiltonian $H_{0}(\boldsymbol{k})+\delta H$ is given by $\chi\xi_{\boldsymbol{k},s}$,
where $\chi=\pm$ denotes the positive or negative energy eigenvalues
and $s=\pm$ denotes the two states in positive or negative energy
band. The free energy density can be obtained as,
\[
\mathcal{F}=\sum_{\eta}\frac{\Delta_{\eta}^{2}}{2g_{\eta}}-\frac{1}{V\beta}\sum_{\chi,s}\sum_{\boldsymbol{k}}\ln\left(1+e^{-\beta\chi\xi_{\boldsymbol{k},s}}\right).
\]
Here $\beta=k_{B}T$ is the product of the Boltzmann constant $k_{B}$
and the temperature $T$. By taking the stationary point of the free
energy density $\partial\mathcal{F}(\Delta_{\eta})/\partial\Delta_{\eta}=0$,
we can obtain the self-consistent equations for $\Delta_{\eta}$ :
\begin{equation}
\Delta_{\eta}=\frac{g_{\eta}}{V}\sum_{s}\sum_{\boldsymbol{k}}\frac{\partial\xi_{\boldsymbol{k},s}}{\partial\Delta_{\eta}}\tanh\frac{\beta\xi_{\boldsymbol{k},s}}{2}.\label{eq:self-consistent}
\end{equation}
The summation over momentum is cut off by the scale $\Lambda/\hbar v$.

For a single SDW order parameter, $\delta H=\Delta_{SDW}\tau_{3}\sigma_{2}$,
the occupied two energy eigenvalues are degenerate with $\xi_{\boldsymbol{k},s}=\sqrt{\Delta_{SDW}^{2}+(\hbar v\boldsymbol{k})^{2}+m^{2}}$.
Eq. (\ref{eq:self-consistent}) can be expressed as
\begin{align*}
\Delta_{SDW} & =2g_{SDW}\int_{0}^{\Lambda/\hbar v}\frac{dkk^{2}}{2\pi^{2}}\frac{\Delta_{SDW}}{\sqrt{(\hbar vk)^{2}+m^{2}+\Delta_{SDW}^{2}}}.\\
 & \times\tanh\frac{\sqrt{(\hbar vk)^{2}+m^{2}+\Delta_{SDW}^{2}}}{2k_{B}T}
\end{align*}
By introducing the dimensionless parameters $\widetilde{g}_{SDW}=\frac{g_{SDW}\Lambda^{2}}{2\pi^{2}(\hbar v)^{3}}$,
$\widetilde{\Delta}_{SDW}=\frac{\Delta_{SDW}}{\Lambda}$ , $\widetilde{m}=m/\Lambda$
and $\widetilde{T}=T/\Lambda$, we obtain the self-consistent equation
for $\widetilde{\Delta}_{SDW}$ in Eq. (\ref{eq:finite_temperature-1}).

\bibliographystyle{apsrev}

\end{document}